\documentstyle[preprint,aps]{revtex}
\begin{document}
\draft
\title{Aspect of Spin-Charge Separation due to Antiferromagentic Spin
Fluctuations in Technically Nested 2D Metals}
\author{Kazumasa MIYAKE and Osamu NARIKIYO}
\address{Department of Material Physics, Faculty of Engineering Science \\
Osaka University, Toyonaka 560, Japan}
\date{\today}
\maketitle
\begin{abstract}
It is shown that the charge susceptibility in nearly half-filled
two-dimensional (2D) metals, with technically nested Fermi surface, shows
anomaly at the wavevector different from that for the spin susceptibility
at low temperatures.  Namely charge degrees of freedom behave there as if
their ``Fermi surface" corresponded to that of ``holes" created in the Mott
insulator.  Such anomaly is
caused by the Aslamazov-Larkin type contribution of 2D antiferromagnetic spin
fluctuations.  This phenomenon gives a possible clue to resolve the paradox
of normal state properties of high-$T_{\rm c}$ cuprates {\it starting}\/ from
the Fermi-liquid fixed point.
\end{abstract}
\pacs{}
\narrowtext
Since the discovery of high-$T_{{\rm c}}$ cuprates, it has been a central
issue how to understand the anomalous properties in the normal state
\cite{anderson1,varma}.  Among them the most exotic one is that the
charge degrees of freedom behave as if they were ``holes" created in the Mott
insulator by doping while the spin degrees of freedom show a response
reflecting the existence of large Fermi surface consistent with the Luttinger
sum rule \cite{fukuyama1}.  This observation seemed to promote
theories which rely on the idea of spin-charge separation in one form or
another \cite{anderson1,fukuyama1,zhang,nagaosa,ioffe,tanamoto,imada}.
Indeed, such theories well explain a considerable part of anomalous
properties.  Furthermore, it has recently been shown by high temperature
expansion \cite{puttika} that the equal time charge correlation function of
2D $t$-$J$ model has a characteristic wavevector corresponding to the Fermi
wavevector of spinless version of the model suggesting the spin-charge
separation to occur in the low temperature limit.  However, it is still
uncertain whether the spin-charge separation can be shown explicitly by
reliable calculations, and some key experiments, such as a systematics of
anomalous behavior of Hall constant \cite{sato,batlogg}, remain to be
unexplained.

On the other hand, the Fermi-liquid theory \cite{landau} in its simple form
fails immediately to explain such anomalies, because the spin and the charge
degrees of freedom are confined there no matter how the electron
correlation is strong as in the heavy fermion systems.
However, some part of anomalies, such as a gross feature of temperature
dependence of the resistivity and the longitudinal NMR relaxation rate, can
be explained by taking into account the spin-fluctuation effect starting from
the Fermi-liquid theory \cite{moriya1,kohno}.  Indeed, in order to understand
such anomalies, it seems still useful to extend the concept of ``adiabatic
continuation" \cite{landau,anderson2} for the Fermi-liquid description and
to take into account the perturbations around the Fermi-liquid fixed point.
Nevertheless, the apparent spin-charge separation aspect observed in the Hall
effect has remained a longstanding unsolved paradox from this point of view.

Recently, we have found \cite{miyake1} that overall feature of the anomaly
concerning the Hall constant can be understood by taking the effect of
antiferromagnetic spin fluctuations of 2D metals with technically nested Fermi
surface into account on the basis of the Fermi-liquid description.
An Aslamazov-Larkin (AL) type contribution \cite{aslamazov} of spin
fluctuations plays a crucial role there.  A purpose of this
paper is to discuss the physical meaning of such a contribution by showing
that the charge susceptibility around the $\Gamma$-point obtains the singular
behavior {\it as the temperature is decreased}\/.  The characteristic
wavevectors are given by difference of incommensurate wavevectors
${\bf Q}_{i}^{*}$ ($i$=1$\sim$4) of dominant
spin fluctuations, because two modes of spin fluctuations with
${\bf Q}\sim{\bf Q}^{*}$ give rise to the singularity at
(${\bf Q}_{i}^{*}-{\bf Q}_{j}^{*}$) which are all located around the
$\Gamma$-point.  Namely, spin and charge degrees of freedom have different
characteristic wavevector, or an aspect of spin-charge separation emerges,
as an effect of mode-coupling between the spin and the charge fluctuations.

It has been shown, in course of microscopic justification of the
self-consistent renormalization (SCR) theory of metallic magnetism
\cite{kawabata,moriya2}, that the charge susceptibility is also subject to
a passive influence of spin fluctuations.  Namely, the Feynman diagram shown
in Fig.\ \ref{fig1} gives the most singular contribution to the charge
susceptibility.  Since it is the property of spin fluctuations that attracted
attentions there, an importance of this AL-type process seems
to have been unrecognized.  The analytic expression
$\kappa^{{\rm AL}}({\bf q})$ for this diagram is given by
\begin{equation}
\kappa^{{\rm AL}}({\bf q})=3T\sum_{\omega_{m}}\sum_{{\bf Q}}
[\gamma_{3}({\bf q},{\bf Q};i\omega_{m})]^{2}
\chi({{\bf Q}},i\omega_{m})\chi({{\bf Q+q}},i\omega_{m})+
\kappa^{{\rm AL}}_{{\rm inc}},
\label{sus1}
\end{equation}
where $\kappa^{{\rm AL}}_{{\rm inc}}$ denotes an incoherent part of AL
contribution which is expected to be non-singular,
$\chi({\bf Q},i\omega_{m})$ denotes a coherent part of the spin-fluctuation
propagator, and $\gamma_{3}$ is the mode-coupling vertex made of three
fermions loop:
\begin{equation}
\gamma_{3}({\bf q},{\bf Q};i\omega_{m})=3T\sum_{\epsilon_{n}}
\sum_{{\bf p}}I^{2}G({\bf p},i\epsilon_{n})G({\bf p+q},i\epsilon_{n})
G({\bf p-Q},i\epsilon_{n}+i\omega_{m}).
\label{gamma}
\end{equation}
The numerical factor 3 in eqs.\ (\ref{sus1}) and (\ref{gamma})
arises from taking trace on spin variables.
Here we have followed the notation of the Hubbard model for simplicity so that
$I$ in eq.\ (\ref{gamma}) denotes the {\it renormalized}\/ on-site repulsion.
However, essentially the same expressions as eqs.\ (\ref{sus1}) and
(\ref{gamma}) are obtained even if we started from the itinerant-localized
duality model \cite{kuramoto} which takes into account key characteristic
of strongly correlated metals, regardless of one-band or multiband model.
The Green function $G$'s in eq.\ (\ref{gamma}) denotes that of coherent part
describing the quasiparticles.

The spin-fluctuation propagator $\chi({\bf Q},i\omega_{m})$ is
determined so as to satisfy a {\it self-consistent} equation of the
mode-coupling
scheme as in Ref.\ \cite{moriya2}.  A difference from the usual case
\cite{moriya1,moriya2} arises if the Fermi surface of quasiparticles is
nested, perfectly or technically.  Such a case has been discussed already
in three dimensions \cite{hasegawa}, where the spin-density-wave long range
order always sets in as temperature is decreased.  We have recently extended
it to 2D case and obtained $\chi({{\bf Q}},i\omega_{m})$ near
the incommensurate antiferromagnetic wavevector ${\bf Q}^{*}_{i}$ in the form
\cite{hasegawa,narikiyo1}
\begin{equation}
\chi({{\bf Q}},i\omega_{m})=\sum_{i}{N_{{\rm F}}\over
\eta+A({\bf Q}-{\bf Q}^{*}_{i})^{2}+C|\omega_{m}|},
\label{chi}
\end{equation}
where $N_{{\rm F}}$ is the renormalized density of states per spin at the
Fermi level, and $A$ and $C$ would be estimated as $A/IN_{{\rm F}}\approx
7\zeta(3)/32\pi^{2}\cdot
v_{{\rm F}}^{2}/T^{2}$, $v_{{\rm F}}$ being the renormalized Fermi velocity,
and $C/IN_{{\rm F}}\approx \pi/8T$ if the circular band were
assumed.  Crudely speaking, $\eta$
in eq.\ (\ref{chi}) shows the temperature dependence like
\begin{equation}
\eta\approx(T+\Theta)/\epsilon_{{\rm F}},
\label{eta}
\end{equation}
where $\epsilon_{{\rm F}}$ is the renormalized Fermi energy, and $\Theta$
the ``Curie-Weiss temperature" parametrizing an extent of deviation from
antiferromagnetic phase boundary.  Deviations from the behavior given by
eq.\ (\ref{eta}) occur both at $T<\Theta$ and $T>T^{*}$, $T^{*}$ being the
characteristic temperature where the character of spin fluctuations crosses
over from itinerant to localized one as discussed elsewhere \cite{narikiyo1}.

Since the dominant contribution of $\kappa^{{\rm AL}}({\bf q})$,
eq.\ (\ref{sus1}), arises from the terms with $\omega_{m}$=$0$ and
${\bf Q}\sim{\bf Q}_{i}^{*}$, it is estimated as
\begin{equation}
\kappa^{{\rm AL}}({\bf q})=3\sum_{i,j}
\gamma_{3}({\bf q},{\bf Q}_{i}^{*};0)\gamma_{3}({\bf q},{\bf Q}_{j}^{*};0)
T\sum_{{\bf Q}}{N_{{\rm F}}\over
\eta+A({\bf Q}-{\bf Q}^{*}_{i})^{2}}
{N_{{\rm F}}\over
\eta+A({\bf Q}+{\bf q}-{\bf Q}^{*}_{j})^{2}}.
\label{sus2}
\end{equation}
Thus $\kappa^{\rm AL}({\bf q})$ is expected to have many peaks
at ${\bf q}=0$,\ $\pm({\bf Q}^{*}_{1}-{\bf Q}^{*}_{2})$,
\ $\pm({\bf Q}^{*}_{1}-{\bf Q}^{*}_{3})$,
\ $\pm({\bf Q}^{*}_{1}-{\bf Q}^{*}_{4})$, where the positions of two peaks of
$\chi({\bf Q},0)$ in eq.\ (\ref{sus2}) coincide each other.  The loci of
$\kappa^{\rm AL}({\bf q})$'s extreme in ${\bf q}$-space are shown
in Fig.\ \ref{fig2} together with those of ${\bf Q}^{*}$'s.  The maximum value
of $\kappa^{\rm AL}({\bf q})$ is reached at ${\bf q}=0$:
\begin{eqnarray}
\kappa^{\rm AL}(0)&=&3\times4[\gamma_{3}(0,{\bf Q}^{*};0)]^{2}
T\sum_{{\bf Q}}{N_{{\rm F}}^{2}\over(\eta+AQ^{2})^{2}}
\label{sus3}\\
&\approx&3{[\gamma_{3}(0,{\bf Q}^{*};0)]^{2}\over\pi A}
\cdot{T\over\eta}\cdot N_{{\rm F}}^{2}.
\label{sus4}
\end{eqnarray}
Here we have left the most singular contribution of ${\bf Q}$-summation
in eq.\ (\ref{sus3}).  The factor 4 in eq.\ (\ref{sus3}) represents number
of independent ${\bf Q}^{*}$ shown in Fig.\ \ref{fig2}.

The factors $\gamma_{3}(0,{\bf Q}^{*};0)$, eq.\ (\ref{gamma}), and $A$ in
eq.\ (\ref{sus4})
are numerically calculated with use of a model dispersion
$\varepsilon_{{\bf k}}$ of renormalized
quasiparticles which has been used to analyze the neutron scattering
experiments of cuprates \cite{tanamoto}:
\begin{equation}
\varepsilon_{{\bf k}}=-2t(\cos k_{x}a+\cos k_{y}a)
-4t^{\prime}\cos k_{x}a\cos k_{y}a
-2t^{\prime\prime}(\cos 2k_{x}a+\cos 2k_{y}a).
\label{dispersion}
\end{equation}
After the summation of $\epsilon_{n}$ is performed in eq.\ (\ref{gamma}),
$\gamma_{3}(0,{\bf Q}^{*};0)$ is expressed as follows:
\begin{equation}
{\gamma_{3}(0,{\bf Q}^{*};0)\over I^{2}}={1\over 2}\sum_{{\bf p}}
{1\over(\xi_{{\bf p}}-\xi_{{\bf p}+{\bf Q}^{*}})}
\biggl[{{\rm th}\displaystyle{\xi_{{\bf p}}\over 2T}-
{\rm th}{\xi_{{\bf p}+{\bf Q}^{*}}\over 2T}
\over(\xi_{{\bf p}}-\xi_{{\bf p}+{\bf Q}^{*}})}-
{1\over 2T}{\rm ch}^{-2}{\xi_{{\bf p}}\over 2T}\biggr],
\label{gamma2}
\end{equation}
where $\xi_{{\bf p}}\equiv\varepsilon_{{\bf p}}-\mu$.  Analytic expression of
$A$ in eq.\ (\ref{chi}) is given by expanding the polarization function with
respect to wavevector around ${\bf Q}^{*}$ as follows:
\begin{eqnarray}
\Pi({\bf Q}^{*}+{\bf q},0)&=&-T\sum_{\epsilon_{n}}\sum_{{\bf p}}
G({\bf p},i\epsilon_{n})G({\bf p}+{\bf Q}^{*}+{\bf q},i\epsilon_{n}) \nonumber
\\
&\approx&\Pi({\bf Q}^{*},0)-{A\over I}{\bf q}^{2}+\cdots.
\label{pi}
\end{eqnarray}
Thus, explicit form of $A$ is given by
\begin{eqnarray}
{A\over I}={1\over2}\sum_{{\bf p}}\biggl\{
{1\over(\xi_{{\bf p}}-\xi_{{\bf p}+{\bf Q}^{*}})^{2}}
& &
\biggl[{{\rm th}\displaystyle{\xi_{{\bf p}}\over 2T}-
{\rm th}{\xi_{{\bf p}+{\bf Q}^{*}}\over 2T}
\over(\xi_{{\bf p}}-\xi_{{\bf p}+{\bf Q}^{*}})}-
{1\over 2T}{\rm ch}^{-2}{\xi_{{\bf p}}\over 2T}\biggr]
\biggl[v^{2}_{x}-{1\over2}(\xi_{{\bf p}}-\xi_{{\bf p}+{\bf Q}^{*}})
{\partial v_{x}\over\partial p_{x}}\biggr] \nonumber \\
& &\ \ \ -{1\over4T^{2}}{1\over(\xi_{{\bf p}}-\xi_{{\bf p}+{\bf Q}^{*}})}
{\rm ch}^{-2}{\xi_{{\bf p}}\over 2T}{\rm th}{\xi_{{\bf p}}\over 2T}
v^{2}_{x}\biggr\}, \label{a}
\end{eqnarray}
where $v_{x}\equiv\partial\varepsilon/\partial p_{x}$.

The result for $[\gamma_{3}(0,{\bf Q}^{*};0)/I^{2}]^{2}/\pi (A/IN_{{\rm F}})$
is shown in Fig.\ \ref{fig3} for parameters, $t^{\prime}/t=-1/6$ and
$t^{\prime\prime}/t=0$, reproducing LSCO-type Fermi surface \cite{tanamoto},
and for the ``hole" concentration $\delta=0.05$ and 0.10.  The energy scale
adopted here is 4$t$, half of the renormalized bandwidth.  It is easily seen
that asymptotic behavior in high temperature region of
$\gamma_{3}(0,{\bf Q}^{*};0)/I^{2}$, eq.\ (\ref{gamma2}), and $A/I$,
eq.\ (\ref{a}), are both proportional to $1/T^{2}$, and so is that of
$[\gamma_{3}(0,{\bf Q}^{*};0)/I^{2}]^{2}/\pi (A/I)$, while the asymptotic
behavior of $A/I$ is not attained yet around $T/\epsilon_{{\rm F}}\sim 0.1$.
The maximum values of $[\gamma_{3}(0,{\bf Q}^{*};0)/I^{2}]^{2}/\pi (A/I)$
reach (2$\sim$3) depending on $\delta$, and so do those of
$\kappa^{{\rm AL}}(0)/N_{{\rm F}}$ because other factors included in
eq.\ (\ref{sus4}) are $T/\eta\sim1$ , $I\sim 1$, and $N_{{\rm F}}\sim1$.
Therefore the singular contribution $\kappa^{{\rm AL}}(0)$, eq.\ (\ref{sus4}),
dominates over the normal part of charge susceptibility
$\kappa^{{\rm N}}(0)\sim 2N_{{\rm F}}$.
It is remarked that the maximum of
$[\gamma_{3}(0,{\bf Q}^{*};0)/I^{2}]^{2}/\pi (A/I)$ for $\delta$=0.05 is
larger than that for $\delta$=0.10.  The reason why the temperature giving
the maximum for $\delta$=0.05 is shifted to higher temperature is probably
due to simple use of a rigid band picuture for the dispersion
(\ref{dispersion}).  It is expected that a real shape of the fully
renormalized Fermi surface approaches the perfect nesting form as the
half-filling is approached, so that the energy scale $h$ characterizing
deviation from perfect nesting decreases making the high temperature
asymptotic behavior survive down to $h$.  Conversely, if there were no nesting
such anomaly of $\kappa^{{\rm AL}}$ would fade away at low temperatures.

Other extreme of $\kappa^{{\rm AL}}({\bf q})$, the loci of which are located
on the ridge surrounding the $\Gamma$-point, are smaller than $\kappa^{{\rm
AL}}(0)$ for geometrical reasons.  The characteristic wavevector of such
extreme
is related with deviation of the incommensurate wavevector ${\bf Q}^{*}$'s
from commensurate one ${\bf Q}_{0}\equiv(\pi/a,\pi/a)$, and is expected to
vanish as $\delta\to 0$.  This is parallel to the finding of Puttika
{\it et al} \cite{puttika} although the details are somewhat different.
The present anomaly can be regarded as 4$k_{{\rm F}}$ singularity, suggested
by Fukuyama {\it et al.}\/ in Ref.\ \cite{fukuyama1}, which is triggered
here by two modes of spin fluctuations with 2$k_{{\rm F}}$ singularity.
This aspect of spin-charge separation is also consistent with the recent
result of Monte Carlo study for the ground state of 2D Hubbard model by
Furukawa and Imada \cite{furukawa} showing that the charge susceptibility
diverges even in {\it intermediate} coupling region ($U$=4$t$) as
$\kappa(0)\propto\delta^{-1}$ while the {\it uniform} spin susceptibility
$\chi(0)$ shows no anomaly at all.  It should be remarked that the dynamical
mass enhancement caluculated by the Gutzwiller treatment is negligible for the
coupling $U/t$=4, i.e.,
$m^{*}/m=(1-\partial\Sigma(\epsilon)/\partial\epsilon)\sim1$ in the
Fermi-liquid sense.

The most divergent corrections to $\kappa^{{\rm AL}}(0)$
are such as those shown in Fig.\ \ref{fig4}.  These are arranged in the
geometric series, so that $\kappa^{{\rm AL}}(0)$ is replaced as
\begin{equation}
\kappa^{{\rm AL}}(0)\rightarrow \kappa^{{\rm AL}}(0)\biggl[1+
{\gamma_{4}\over4\pi A}\cdot{T\over \eta}\biggr]^{-1},
\label{higher}
\end{equation}
where $\gamma_{4}$ is low energy limit of the square vertex, the mode-coupling
constant of spin fluctuations
$\chi({\bf Q}^{*})$'s.  High temperature asymptotic behavior of $\gamma_{4}$
is given as $\gamma_{4}\approx7\zeta(3)N_{{\rm F}}/8\pi^{2}T^{2}$, the same
as that of $A$, for the circular band.  Therefore, the singularity of the
right hand side in the low temperature region is expected to remain the same
as $\kappa^{{\rm AL}}(0)$.

It is noted that a possible van Hove singularity in the polarization function
$\Pi(0,0)$, eq.\ (\ref{pi}), is easily suppressed by taking only RPA
diagrams into account: $\kappa^{{\rm RPA}}(0)=2\Pi(0,0)/[1+I\Pi(0,0)]$.

In 3D case, the present anomaly for $\kappa^{{\rm AL}}$ is weakened
considerably because of phase space reason.  Indeed, ${\bf Q}$-summation
in eq.\ (\ref{sus3}) results in a factor $A^{-3/2}\cdot\ln(1/\eta)$ instead
of $A^{-1}\cdot\eta^{-1}$ leaving a weak singularity as
$\kappa^{{\rm AL}}(0)\sim\ln(1/\eta)$.  Here we have used the fact that the
singularity of $\gamma_{3}(0,{\bf Q}^{*};0)/I^{2}$ and $A/I$ are both
$1/T^{2}$ the same as in 2D.

In conclusion, it has been shown that the charge susceptibility in nearly
half-filled 2D metals has pronounced singularity as decreasing temperature
at the wavevector corresponding to two modes of spin fluctuations if the
Fermi surface is technically nested.  As a result the charge degrees of
freedom show response in the low temperature region as if the ``Fermi surface"
were for the ``holes" created in the Mott insulator.  Such anomaly is caused
by the AL-type contribution of 2D antiferromagnetic spin fluctuations.
This phenomenon gives a possible clue to resolve the paradox
of normal state properties of high-$T_{\rm c}$ cuprates {\it starting}\/
from the Fermi-liquid fixed point.

One of us (K. M.) acknowledges N. Nagaosa for conversation reminding him of
an intimate relation of the present work to that of Ref.\ \cite{furukawa}.
This work is supported by the Grant-in-Aid for General Scientific Research
(04640369), and Monbusho International Scientific Program, Joint Research
``Magnetism and Superconductivity in Highly Correlated Systems" (03044037),
and the Grant-in-Aid for Scientific Research on Priority Areas ``Science of
High Temperature Superconductivity" (04240103) of Ministry of Education,
Science and Culture.

\begin{figure}
\caption{Diagram for the most singular contribution of spin fluctuations
to the charge susceptibility $\kappa^{{\rm AL}}$.  Wavy lines denote the
spin-fluctuation propagator, eq.\ (3), and triangle the vertex $\gamma_{3}$,
eq.\ (2).}
\label{fig1}
\end{figure}

\begin{figure}
\caption{(a) Position of incommensurate antiferromagnetic wavevector
${\bf Q}^{*}$ shown by closed circles and wavevectors for the extremes of
$\kappa^{{\rm AL}}$ shown by arrows.  (b) Locus of extreme
of $\kappa^{{\rm AL}}({\bf q})$ shown by closed circle the area of which
represents a degree of singularity.}
\label{fig2}
\end{figure}

\begin{figure}
\caption{Numerical results of
$[\gamma_{3}(0,{\bf Q}^{*};0)/I^{2}]^{2}/\pi (A/I)$ in
eq.\ (7) for LSCO-type dispersion. }
\label{fig3}
\end{figure}

\begin{figure}
\caption{Series of diagram for the most dominant cerrections to
$\kappa^{{\rm AL}}(0)$.  \ \ \ \ \ \ \ \ \ \ \ \ \ \ \ \ \ \ \ \ \ \ \ \ }
\label{fig4}
\end{figure}

\end{document}